\documentclass[runningheads,a4paper]{llncs}

\usepackage{amssymb}
\usepackage{graphicx}

\usepackage{url}
\urldef{\mailsa}\path|{alfred.hofmann, ursula.barth, ingrid.haas, frank.holzwarth,|
\urldef{\mailsb}\path|anna.kramer, leonie.kunz, christine.reiss, nicole.sator,|
\urldef{\mailsc}\path|erika.siebert-cole, peter.strasser, lncs}@springer.com|    
\newcommand{\keywords}[1]{\par\addvspace\baselineskip
\noindent\keywordname\enspace\ignorespaces#1}

\begin{document}

\mainmatter  

\title{Tracking accuracy evaluation of electromagnetic sensor-based colonoscope tracking method}

\titlerunning{Tracking accuracy evaluation of colonoscope tracking method}

%
%
\author{Masahiro Oda \inst{1}
\and Hiroaki Kondo \inst{1} \and Takayuki Kitasaka \inst{2} \and Kazuhiro Furukawa \inst{3} \and Ryoji Miyahara \inst{3} \and Yoshiki Hirooka \inst{4} \and Hidemi Goto \inst{3} \and Nassir Navab \inst{5} \and Kensaku Mori \inst{6,1}}
%
\authorrunning{Masahiro Oda et al.}

\institute{Graduate School of Information Science, Nagoya University,\\
Furo-cho, Chikusa-ku, Nagoya, Aichi, 464-8603, Japan\\
\email{moda@mori.m.is.nagoya-u.ac.jp}\\
\and
School of Information Science, Aichi Institute of Technology,
\and
Department of Gastroenterology and Hepatology, Nagoya University Graduate School of Medicine,
\and
Department of Endoscopy, Nagoya University Hospital,
\and
Technische Universit\"{a}t M\"{u}nchen,
\and
Strategy Office, Information and Communications, Nagoya University,
\\
\url{http://www.mori.m.is.nagoya-u.ac.jp/~moda/index-e.html}
}

%
%

\toctitle{Lecture Notes in Computer Science}
\tocauthor{Authors' Instructions}
\maketitle

\begin{abstract}
This paper reports a detailed evaluation results of a colonoscope tracking method. A colonoscope tracking method utilizing electromagnetic sensors and a CT volume has been proposed. Tracking accuracy of this method was evaluated by using a colon phantom. In the previously proposed paper, tracking errors were measured only at six points on the colon phantom for the accuracy evaluation. The point number is not enough to evaluate relationships between the tracking errors and positions in the colon. In this paper, we evaluated the colonoscope tracking method based on more detailed measurement results of the tracking errors. We measured tracking errors at 52 points on the colon phantom and visualized magnitudes of the tracking errors. From our experiments, tracking errors in the ascending and descending colons were enough small to perform colonoscope navigations. However, tracking errors in the transverse and descending colons were large due to colon deformations.
\end{abstract}

\section{Introduction}

Colonoscopy is conventionally performed as a colon diagnosis or inspection method.
However, colonoscopy may cause discomfort for patients while diagnosis.
Also, colonoscopy has a risk of complication including perforations of the colon.
Success of colon diagnosis under a colonoscope is heavily depends on physician's skill.

CT colonography (CTC) is a colon diagnosis method that reduces discomfort and a risk of complication on patients.
Diagnosis is performed by using CT images of a patient in CTC.
Some computer aided diagnosis (CAD) systems for CTC are commercially available.
These systems commonly display 2D or 3D or unfolded views of the colon for observation.
Physicians diagnose the colon to find polyps or cancers.

Polyps or cancers of the early stages found by CTC CAD systems can be removed in colonoscopic polypectomy.
Colonoscopic polypectomy is a surgery to remove polyps or cancers.
In colonoscopic examinations including colonoscopic polypectomy, a physician controls a colonoscope based on his/her experience.
Experienced physicians remove polyps or cancers minimizing patient discomfort.
However, colonoscopic examinations performed by inexperienced physicians may painful for patients.
Utilization of a navigation system for colonoscope is one solution for such problem.
Colonoscope navigation systems indicate positions of the colonoscope tip and targets such as polyp positions while performing colonoscopic examinations.
Colonoscope navigation systems can be used to reduce overlooking of polyps and to assist inexperienced physicians.
Conventionally, information obtained from CT volumes of patients is utilized only for the diagnosis stage including diagnosis using CTC CAD systems.
Information obtained from CT volumes contains the polyp positions and the colon shapes.
Such information is useful for the treatment stage including colonoscopic examinations.
The colon shapes obtained from CT volumes can be used as maps of the colon in colonoscope navigation systems.
Also, the polyp positions can be used as target point in navigations.
A colonoscope will be navigated to polyp positions while performing colonoscope examinations by utilizing information obtained from CT volumes of patients.

To achieve colonoscope navigation systems, tracking method for colonoscope is required.
Tracking methods of endoscopes including colonoscope have been proposed by many research groups, which estimate an endoscope tip position in the organs.
For bronchoscope tracking, image-based\cite{Peters08,Deligianni05,Rai08,Deguchi09} and sensor-based\cite{Gildea06,Schwarz06}  tracking methods were reported.
Colonoscope tracking is difficult compared to the tracking for other hollow organs because the colon greatly deforms during an insertion of the colonoscope.
Liu et al.\cite{Liu13} tried to estimate colonoscope tip movements from the optical flow of colonoscope videos.
This method can track a colonoscope tip without using additional equipments for the tracking.
However, tracking using colonoscope videos is easily fails when unclear video frames appear.
Unclear video frames frequently appear in colonoscope video because fluid, feces, and bubbles exist in the colon.
The colonoscope tip touches the wall of the colon many times while colonoscope examinations.
It causes black video frames that make interruptions of tracking.
A colonoscope shape tracking system, the Olympus ScopeGuide (UPD-3), is commercially available.
The system detects colonoscope shape in the colon using electromagnetic (EM) position sensors.
Clinical reports about utilization of the system in colonoscopy have been reported\cite{Fukuzawa15}.
The system just displays the shape of the colonoscope without combining CT volumes or CTC information.
Oda et al.\cite{Oda13,Oda14} and Kondo et al.\cite{Kondo14} proposed colonoscope tracking method using EM position sensors with combination of CT volumes.
They attach EM sensors to colonoscope to obtain the colonoscope shape.
They obtain the colon shape of a patient from a CT volume.
Correspondences between the colonoscope and colon shapes by applying two steps correspondence finding processes.
Based on the correspondences, they find a point in the CT volume which corresponds to the colonoscope tip position.
These methods can track the colonoscope tip position even if the viewing fields of the colonoscope are not clear.
In their tracking error evaluations, they measured tracking errors only at six points in a colon phantom.
Behaviors of the colon deformation are differ according to position.
Therefore, tracking errors should evaluated at many positions in the colon.

In this paper, we perform detailed evaluations of the tracking errors of the colonoscope tracking method using EM sensors\cite{Kondo14}.
We measured tracking errors at 52 points in a colon phantom by using the tracking method.
Based on the measurement results, we discuss relationships between the colon deformations and the tracking errors.

In the section \ref{sec:method}, we briefly introduce the colonoscope tracking method proposed in the reference\cite{Kondo14}.
Experimental results including tracking error measurement results are shown in the section \ref{sec:experiments}.
Discussion about the experimental results are described in the section \ref{sec:discussion}.

\section{Method
\label{sec:method}}

\subsection{Colon centerline and colonoscope line generation}

A colon centerline is obtained from a CT volume.
We use a region growing method to extract a colon region from the CT volume.
A thinning and a line smoothing processes are applied to generate a colon centerline.

A colonoscope line that represents the colonoscope shape is obtained by using EM position sensors.
We insert an Aurora 5/6 DOF Shape Tool Type 1 (NDI) to the colonoscope working channel.
The shape tool gives positions and directions of the colonoscope at seven points.
The colonoscope line is calculated by applying the Hermite spline interpolation to positions and directions measured by the shape tool.

\subsection{Coordinate system registration}

We generate a modified colon centerline that simulates the shape of the colon while an insertion of the colonoscope.
This process is required because the colon largely deformed while colonoscope insertions.
To generate the modified colon centerline, we detect sections on the colon centerline that corresponds the transverse and sigmoid colons.
The transverse and sigmoid colon sections on the colon centerline are replaced with straight line sections.
The transverse and sigmoid colon sections are identified based on the positions and the shape of the colon centerline.

We register the CT and sensor coordinate systems by using the ICP algorithm\cite{Besl92}.
The ICP algorithm finds a rigid transformation matrix that minimizes the Euclidean distance between the modified colon centerline and colonoscope line.
The colonoscope line is transformed to the CT coordinate system by using the rigid transformation matrix.

\subsection{Colonoscope tip position finding}

We find correspondences between each point on the colon centerline and colonoscope line.
This process consists of two steps including a landmark-based coarse correspondence finding and a length-based fine correspondence finding.
The landmark-based coarse correspondence finding process finds corresponding point pairs on the colon centerline and colonoscope line at five anatomical landmarks.
The five anatomical landmarks are detected based on their positions and the shape of the colon.
After performing the coarse correspondence finding, the length-based fine correspondence finding is applied.
The length-based fine correspondence finding process finds corresponding point pairs on the colon centerline and colonoscope line at all points on them.
This process finds correspondences by using lengths along the lines.
Finally, a point on the colon centerline that corresponds to the tip of the colonoscope line is defined as the colonoscope tip position in the CT coordinate system.

\section{Experiments
\label{sec:experiments}}

\begin{figure}[tb]
\begin{center}
\begin{tabular}{cc}
\includegraphics[width=0.53\textwidth]{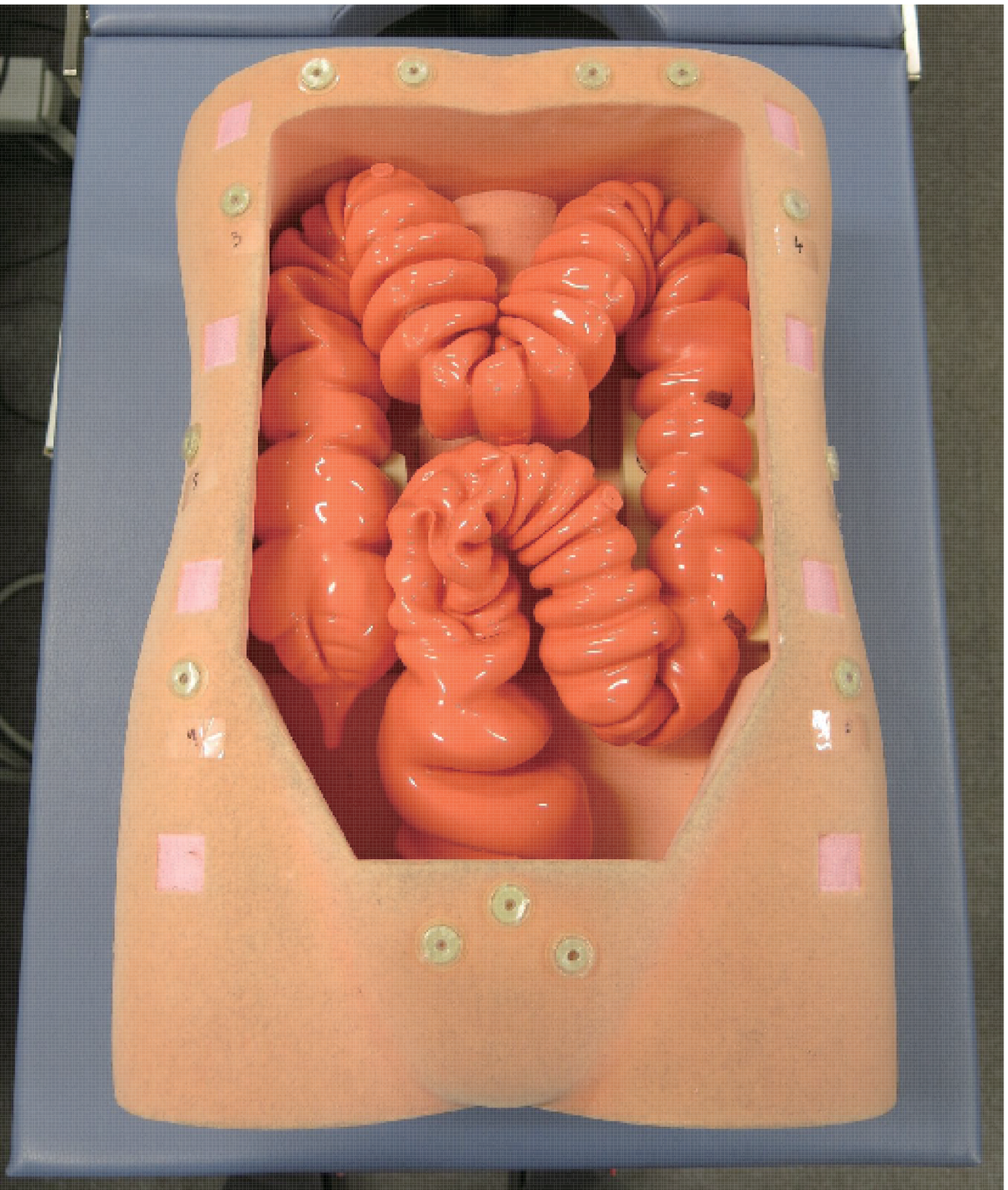}
&
\includegraphics[width=0.43\textwidth]{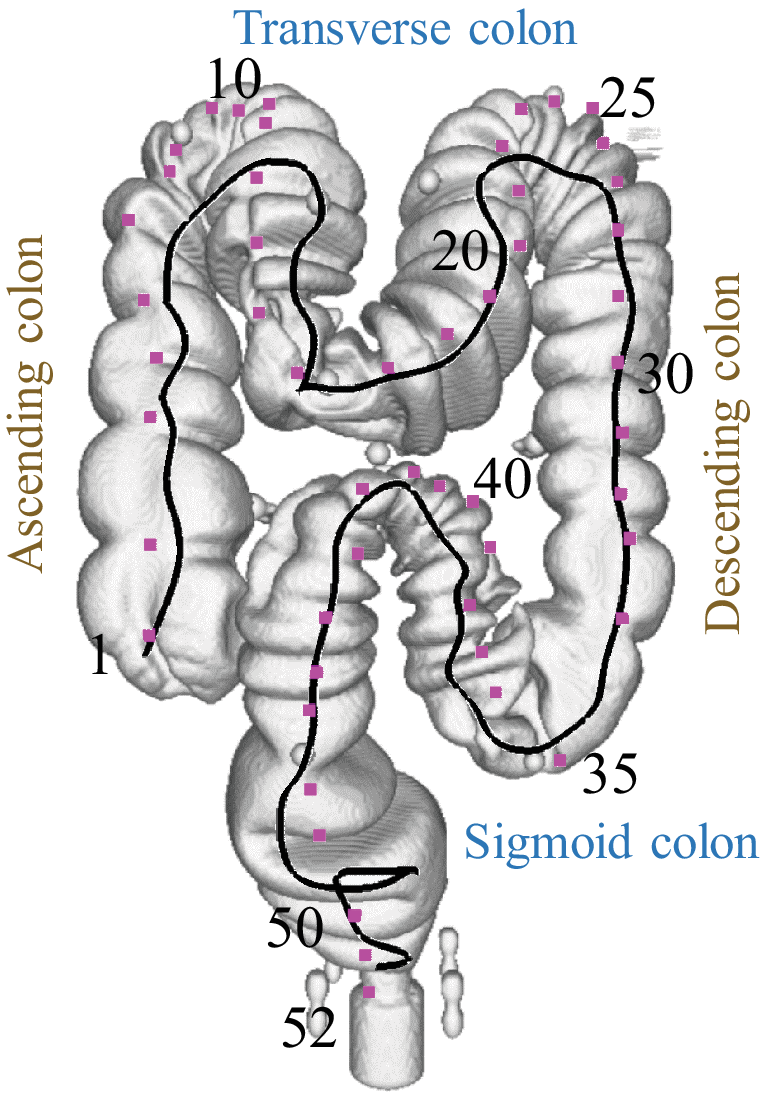}\\
(a) & (b)
\end{tabular}
\caption{(a) Colon phantom. (b) Positions of EPs placed on the surface of the colon phantom. Purple points are EPs and numbers are indices of EPs. Black line is colon centerline.}
\label{fig:colonphantom}
\end{center}
\end{figure}

The proposed method was evaluated in experiments using a colon phantom.
A colon phantom (KOKEN colonoscopy training model type I-B) (Fig. \ref{fig:colonphantom} (a)) and its CT volume are utilized in our experiments.
We used a colonoscope (Olympus CF-Q260AI).
An Aurora 5/6 DOF Shape Tool Type 1 (NDI) is inserted to the working channel of the colonoscope.
The Aurora 5/6 DOF Shape Tool Type 1 has seven EM sensors.

In colonoscopic examinations, physicians observe the colon by using a colonoscope while pulling back the colonoscope after insertion up to the cecum.
To simulate this situation, we inserted the colonoscope up to the cecum of the colon phantom before performing the colonoscope tracking.
After the insertion, we started the colonoscope tracking and measured tracking errors while pulling back the colonoscope.
Definition of the tracking error is described below.

We evaluated the performance of the proposed method by using a tracking error.
We defined evaluation points (EPs) on the surface of the colon phantom.
Points on the colon phantom surface which have characteristic shapes (such as parts of the haustral folds or taeniae coli) were selected as EPs.
The EPs are visually identifiable from both of the colon phantom and its CT volume.
Positions of the EPs and indices of them are shown in the Fig. \ref{fig:colonphantom} (b).
The position of each EP was projected to the closest point on the colon centerline.
The tracking error is the length along the colon centerline between an estimated position of the colonoscope tip (estimated by the colonoscope tracking method) and a position of a projected EP when the real colonoscope tip comes to the closest position to the marker.

\begin{figure}[tb]
\begin{center}
\includegraphics[width=1.0\textwidth]{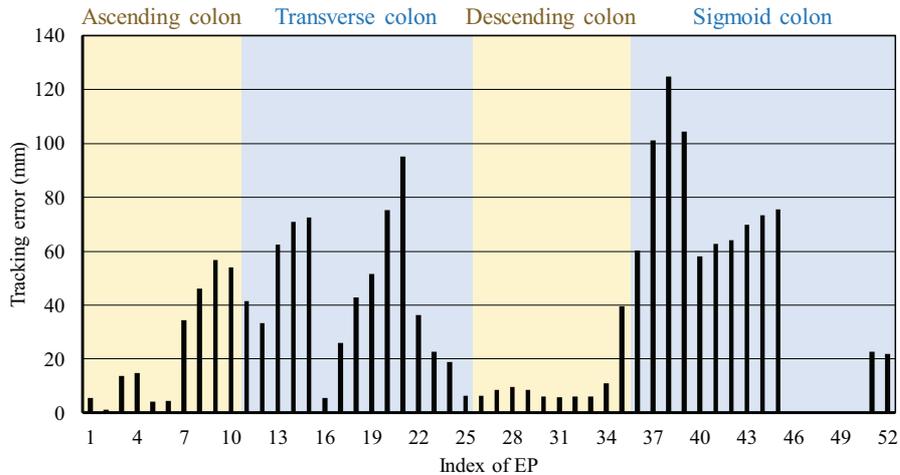}
\caption{Average tracking errors at EPs  from three trials. Areas correspond to the ascending, transverse, descending, and sigmoid colons are indicated by yellow and blue colors. Average tracking errors of EPs having indices from 46 to 50 are not shown here because measurements were failed.}
\label{fig:trackingerror_graph}
\end{center}
\end{figure}

\begin{figure}[tb]
\begin{center}
\includegraphics[width=0.6\textwidth]{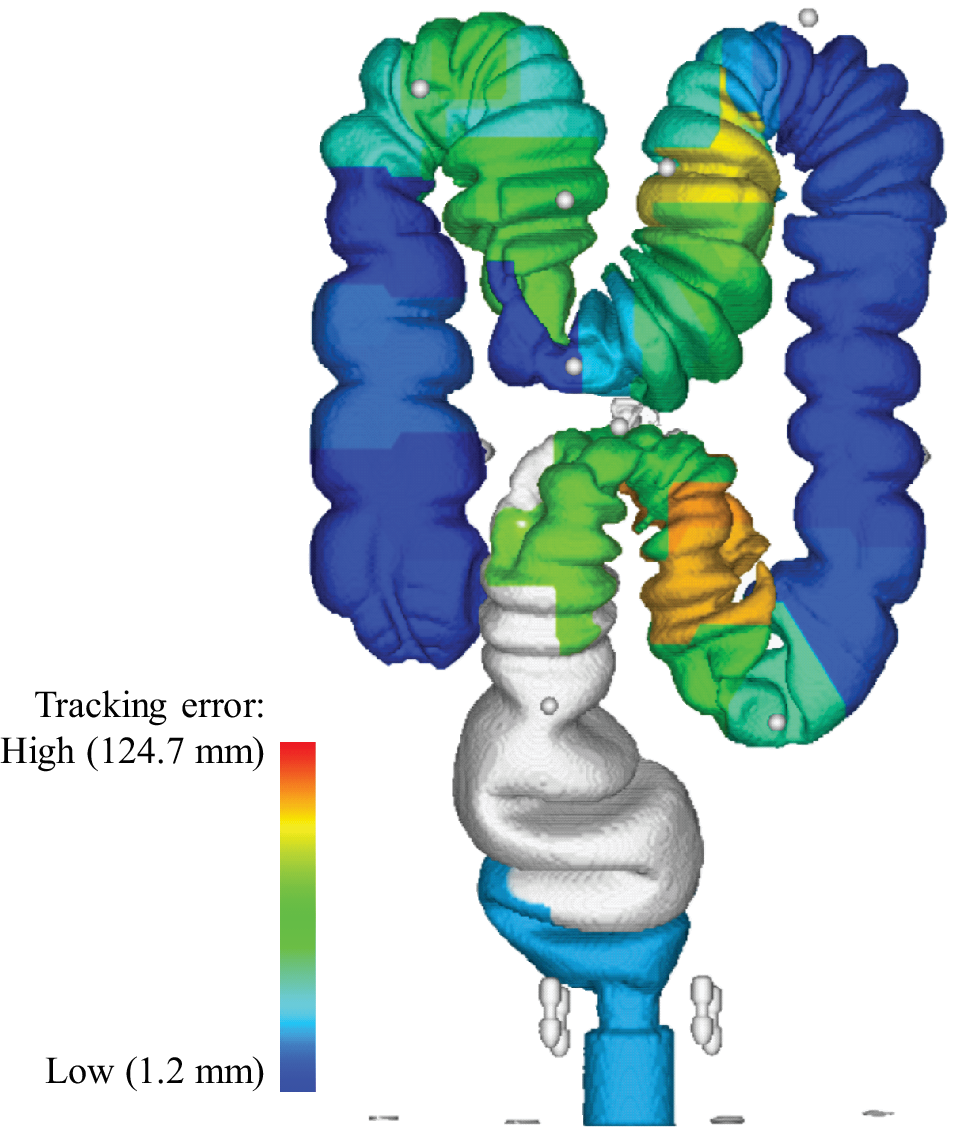}
\caption{Average tracking errors at EPs indicated by colors on the colon phantom. Blue and red colors indicate small and large average tracking errors. White color means measurement was failed at the region due to deformations of the colon phantom while pulling the colonoscope. This image is rendered from a CT volume of the colon phantom.}
\label{fig:trackingerror_color}
\end{center}
\end{figure}

In the reference\cite{Kondo14}, tracking errors were measured at six EPs.
We performed a detailed evaluation of the colonoscope tracking method by using 52 EPs.
We measured tracking errors of colonoscope insertions in three trials.
Figure \ref{fig:trackingerror_graph} shows the average tracking errors at each EPs.
The segments of the colon (ascending, transverse, descending, and sigmoid colons) are also shown in this figure.
Measurements were failed at the EPs of indices from 46 to 50 because we could not find these EPs due to large deformations of the colon phantom while pulling the colonoscope.
We showed the average tracking errors by using colors on the colon phantom as Fig. \ref{fig:trackingerror_color}.
In this figure, blue and red colors indicate small and large average tracking errors.
Small tracking errors were obtained in the ascending and descending colons.
Large tracking errors were obtained in the transverse and sigmoid colons.


\section{Discussion
\label{sec:discussion}}

From the experimental results, relations between regions in the colon and tracking errors were clearly shown.
We measured tracking errors at 52 EPs.
The number of EPs was significantly larger than the reference\cite{Kondo14}.
Our result is useful to investigate causes of the tracking errors.

The tracking errors were quite small in the ascending and descending colons.
Most of the tracking errors in the regions were smaller than 40 mm.
Tracking errors in the regions were enough small to perform colonoscope navigations.
A physician who specializes in gastroenterology said that tracking errors smaller than 50 mm are acceptable for navigations of the colonoscope tip to polyps.
If a polyp comes to a position near the colonoscope tip (about 50 mm or closer), it is observable from the colonoscope camera.
The tracking method is applicable for colon navigations to find polyps in the ascending and descending colons.
The tracking errors were small in the ascending and descending colons because these regions not deform largely.
The ascending and descending colons are fixed to the other tissues.
It makes small tracking errors in these regions.

Unlike the ascending and descending colons, the transverse and sigmoid colons largely moves in the abdominal cavity.
The transverse and sigmoid colons largely deform during colonoscope insertions.
The shapes of the transverse and sigmoid colons during the colonoscope insertions are nearly straight.
It caused the large tracking errors in the transverse and sigmoid colons.
Estimation method of colon deformations during colonoscope insertions is required to reduce tracking errors.

\section{Conclusions}

This paper reported a detailed evaluation results of the tracking errors of the colonoscope tracking method.
The colonoscope tracking method estimates the colonoscope tip position in the colon by using EM sensors and a CT volume.
We measured average tracking errors at 52 points in a colon phantom by using the tracking method.
Three trials of measurements were performed.
The average tracking errors in the ascending and descending colons were enough small to perform colonoscope navigations.
However, the average tracking errors in the transverse and sigmoid colons were large due to deformations of the colon.

\subsubsection*{Acknowledgments.}
Parts of this research were supported by the MEXT, the JSPS KAKENHI Grant Numbers 24700494, 25242047, 26108006, 26560255, and the Kayamori Foundation of Informational Science Advancement.

\end{document}